\documentclass[prd,12pt,a4paper,nofootinbib,showpacs]{revtex4}
\usepackage[left=2cm,right=2cm,top=2.5cm,bottom=2.5cm]{geometry}

\usepackage[frenchb,english]{babel}
\usepackage[applemac]{inputenc}
\usepackage{enumerate}
\usepackage{amsmath}
\usepackage{amsthm}
\usepackage{amssymb}
\usepackage{graphicx}
\usepackage{floatflt}
\usepackage{fancybox}
\usepackage{stmaryrd}
\usepackage{appendix} 
\usepackage{enumitem} 

\usepackage{hyperref} 

\usepackage{color}

\newcommand{\Ai}{\textrm{Ai}}

\newcommand{\Om}{\Omega}

\newcommand{\lam}{\lambda}

\newcommand{\eps}{\epsilon}
\newcommand{\Lam}{\Lambda}

\renewcommand{\P}{p_a}
\newcommand{\eff}{{\rm eff}}
\newcommand{\ma}{{\rm mat}}

\newcommand{\sign}{\textrm{sign}}

\newcommand{\p}{\partial}

\newcommand{\M}{\mathcal M_{if}}
\newcommand{\el}{{\rm el}}

\newcommand{\be} {\begin{equation}}
\newcommand{\ee} {\end{equation}}
\newcommand{\bsub}{\begin{subequations}}
\newcommand{\esub}{\end{subequations}}
\newcommand{\bea}{\begin{eqnarray}}
\newcommand{\eea}{\end{eqnarray}}
\newcommand{\bi} {\begin{itemize}}
\newcommand{\ei} {\end{itemize}}
\newcommand{\ben} {\begin{enumerate}}
\newcommand{\een} {\end{enumerate}}
\newcommand{\bmat} {\begin{pmatrix}}
\newcommand{\emat} {\end{pmatrix}} 
\newcommand{\bal} {\begin{aligned}}
\newcommand{\eal} {\end{aligned}}
\newcommand{\btab}{\begin{tabular}}
\newcommand{\etab}{\end{tabular}}

\newcommand{\Sec}[1]{Sec.~\ref{#1}}

\newcommand{\eq}[1]{Eq.~\eqref{#1}}

\begin{document}
\selectlanguage{english}

\title{Unitary and non-unitary transitions around a cosmological bounce}

\author{Antonin Coutant}
\email{antonin.coutant@aei.mpg.de}
\affiliation{Max Planck Institute for Gravitational Physics, Albert Einstein Institute, Am Muhlenberg 1, 14476 Golm, Germany, EU}

\date{\today}

\begin{abstract}
In this work, we investigate the notion of time and unitarity in the vicinity of a bounce in quantum cosmology, that is, a turning point for the scale factor. Because WKB methods drastically fail near a turning point, the scale factor cannot play the role of time in such scenarios. We overcome this difficulty by studying the dynamics of matter transitions when using its conjugate momentum as a time. We find precise conditions so as to recover unitarity, and hence, a consistent notion of probability. We then compute transitions in a concrete example, extract the specific feature of a bounce and argue about the necessity of the conjugate momentum representation to go beyond the background field approximation. Our analysis is also equally relevant for a closed Universe undergoing a recollapsing phase.
\end{abstract}


\hfill \small AEI-2014-024

\maketitle


\newpage
\section{Introduction}
A well-known feature of diffeomorphism invariant theories is the absence of time from the quantum dynamical equations~\cite{DeWitt67a}. In non-perturbative quantum gravity, all degrees of freedom have their conjugate momentum contributing to the Hamiltonian quadratically. For this reason, none of them can be used as a ``perfect clock'' so as to recover a Schrödinger type of equation~\cite{UnruhWald89}. In the absence of a time parameter, it is unclear how to extract probabilities from the solutions of the Wheeler-DeWitt equation, as this parameter plays a crucial and specific role in usual quantum theories~\cite{Kuchar91,Isham92,CarterHartle}. This raises the question of the interpretation of the wave function in the Wheeler-DeWitt equation. 

In order to make connection with the usual quantum formalism, one degree of freedom, say $\lam$, may be used to play the role of ``time''. In the lines of~\cite{Halliwell84,Brout88,Brout98}, we adopt here the point of view of matter. In other words, the parameter $\lam$ is used to describe the rate of processes. This approach corresponds to ``choosing time after quantization'' in~\cite{Isham92}. We then define a Hilbert space at fixed values of $\lam$. To implement the usual interpretation of the wave function, we must find a positive definite scalar product conserved through the evolution in $\lam$. To see this, we may consider the ``evolution operator'', that relates the wave function at $\lam = \lam_1$ to the one at $\lam = \lam_2$, 
\be
|\psi(\lam_2) \rangle = U(\lam_2,\lam_1) \cdot |\psi(\lam_1) \rangle. 
\ee
If this operator is unitary with respect to some (positive definite) scalar product, then a consistent probability interpretation can be implemented. Note that this is close to the historical investigation of Born that led him to formulate the probabilistic interpretation of quantum mechanics~\cite{Born26}. In~\cite{Vilenkin88}, Vilenkin also followed this line of thought. Exploiting the conserved current of the Wheeler-DeWitt equation to build probabilities, he established that if one gravitational degree of freedom is semiclassical, the evolution operator becomes approximately unitary. In isotropic cosmological models, this amounts to choosing the scale factor $a$ as a time parameter. Later, Massar and Parentani obtained precise conditions to obtain a unitary evolution when $a$ plays the role of time~\cite{Massar97,Massar98}. Interestingly, they showed that a unitary regime is reached \emph{before} this degree of freedom becomes completely classical. In other words, there exist a regime where the evolution is unitary, but there is still no consistent notion of background metric and therefore no time in a Schrödinger sense. 

Unfortunately, this interpretation fails in many physically relevant situations. In particular, it cannot be applied near a \emph{turning point} for $a$. This failure is equally present in other approaches involving the WKB method or appealing to a ``WKB time'' (see~\cite{Isham92} and references therein). In this work, we propose to reinvestigate the question by using $\P$, the conjugate momentum of $a$, as the time parameter. This allows us to go beyond the adiabatic approximation~\cite{Hajicek86,KeskiVakkuri96}, and analyze quantum transitions of matter degrees of freedom. Near a turning point, while the transitions seemed highly non-unitary with the $a$-time, we find a well-defined unitary regime when using $\P$ as a time. The main condition to obtain such a regime, is that semiclassical trajectories corresponding with different monoticities of $\P$ in time must decouple. Additionally, we also find that unitarity is recovered before the background metric approximation. In this intermediate regime, unitarity is a well defined concept, and backreaction is still implemented at a quantum level. We also point out that this probability interpretation, which uses a scalar product conserved in $\P$, is not equivalent to the conserved current of Vilenkin~\cite{Vilenkin88}. This result agrees with the more general statement that the notion of unitarity is intrinsically related to the choice of a time parameter, a conclusion also reached in different approaches~\cite{UnruhWald89,Marolf94,Marolf09,Bojowald10,Bojowald10b,Hohn11}. In the first section, we present our cosmological model. In the second section, we derive the dynamical equation for matter transitions in $\P$-time, and obtain precise conditions to recover unitarity. We point out that this second part is derived in a much more general framework than the presented model. In the last section, we consider our model and investigate matter transitions in the intermediate regime, i.e., unitary transitions without background metric, in the vicinity of a turning point. 

\section{Cosmological model}
\subsection{Cosmological bounce}
In this work, we consider a homogeneous and isotropic Universe. The classical space-time is described by a FLRW metric 
\be
ds^2 = N(\lam)^2 d\lam^2 - a(\lam)^2 d\Om_{K}^2. \label{FLRW}
\ee
$\lam$ is a general time coordinate, and $d\Om_{K}^2$ is the spatial metric of constant curvature $K = 0, \pm 1$. Under such a symmetry reduction, the only gravitational degree of freedom is the scale factor $a$. The lapse function $N$ generates the Hamiltonian constraint, that guarantees that the theory is invariant under time reparametrization~\cite{Wald}. In minisuperspace, this constraint is the only equation of motion. The dynamics of gravity and matter fields is given by the action\footnote{To lighten the notations, we have redefined the gravitational constant as $G = 4\pi G_{\rm N}/3$, where $G_{\rm N}$ is the standard Newton constant. This is equivalent to add a numerical factor to the metric \emph{ansatz} \eqref{FLRW}, as done e.g. in~\cite{CarterHartle}.} 
\be
\mathcal S = \frac1{2G} \int \left(- \frac{a \dot a^2}N - \frac{V(a)N}{a} \right) d\lam + \mathcal S_\ma. \label{action}
\ee
$V(a)$ is the ``super potential'' of the Wheeler-DeWitt equation~\cite{DeWitt67a}. In minisuperspace with a cosmological constant $\Lam$, it reads $V(a) = - K a^2 + \Lam a^4/3$. In the following, it is replaced by an effective potential $V_\eff(a)$. This effective potential takes into account energy contributions from matter fields at equilibrium (e.g.,  in a radiation dominated Universe $V_\eff^{\rm rad}(a) = 2G \rho_{\rm rad}^0$, with $\rho_{\rm rad}^0$ an $a$-independent reference radiation density) as well as quantum gravity effects, as in loop quantum cosmology~\cite{Ashtekar11,Bojowald12}. In this paper, we do not choose nor justify a specific form. We simply postulate that there is a turning point for the variable $a$, and study the consequences on the dynamics of matter fields. 

A turning point is a transition from a contracting to an expanding phase (or vice-versa), and arise in cosmological scenarios where the Universe undergoes a ``bounce''. This occurs for example in loop quantum cosmology~\cite{Ashtekar11}, but also in the presence of ``exotic matter'', as for instance in~\cite{Bekenstein75}. In these scenarios, when the Universe becomes too dense, deviations from the standard Big Bang model violate the positive energy conditions and induce a bounce. A turning point for $a$ also occurs for a closed Universe that undergoes a recollapsing phase, i.e., after the phase of expansion, the gravitational attraction finally overcome, and the Universe starts contracting toward a ``Big Crunch'' singularity. Even though our analysis equally applies to the latter case, in the discussions, we shall have in mind the first case: the bounce is generated by quantum gravity or other microscopic effects at high density. Independently of its origin, our main concerns are its consequences on the physics of matter fields. Note also that strictly speaking, our analysis is valid for scenarios with a single turning point, and no turning point for the variable $\P$. This excludes the full description of a tunneling process or cyclic Universes. Nevertheless, our results should stay valid in a neighborhood (not necessarily ``small'') of any turning point for $a$. 

\subsection{Field content}
\label{content_Sec}
By canonically quantizing the action of \eq{action}, we obtain the minisuperspace Wheeler-DeWitt equation~\cite{CarterHartle,Isham92}. It is well-known that this quantization leads to ordering ambiguities~\cite{Isham92,Kiefer13}. Here, we do not address such a delicate issue, and choose the most convenient ordering. Note however that changing the ordering can be absorbed into a redefinition of the wave function and by changing the effective potential. Hence our findings stay valid in the general case. When coupling gravity to matter, the Wheeler-DeWitt equation reads 
\be
\left[ G^2 \p_a^2 + V_\eff(a) + 2Ga \hat H_\ma(a) \right] \Psi(a,\phi_i) = 0,  \label{Wheeler-DeWittEq}
\ee
where $H_\ma(a)$ is the Hamiltonian of matter degrees of freedom. In what follows, we shall focus on a concrete example: a Universe filled with conformal radiation and heavy fields~\cite{Brout98,Parentani96,Parentani96b}. More precisely, we first consider a conformal massless scalar field\footnote{The assumption of a conformal coupling ensures that the field evolves adiabatically for \emph{all} values of $k$ while the Universe expands (without interactions). One could also relax this assumption by considering $k$ large enough so that $(\dot a / k)^2 \ll 1$, similarly to massive fields. \label{massless_ftn}} 
$\phi$ of fixed momentum of norm $k > 0$. In addition, we add two real massive fields $\psi_\mu$ and $\psi_M$ of respective masses $\mu < M$. Those two fields are considered to be heavy enough so that pair creation due to the Universe expansion may be neglected. As we show in App.~\ref{Adiab_App}, this requires $(\dot a / a \mu)^2 \ll 1$, where the derivative is taken with respect to comoving time~\cite{BirrellDavies} (similarly for $\psi_M$). Assuming this, after a rescaling of the fields (see App.~\ref{Adiab_App} for details), the free matter Hamiltonian reads
\be
\hat H_0 = M d_M^\dagger d_M + \mu d_\mu^\dagger d_\mu + \frac{k}{a} d_\phi^\dagger d_\phi .
\ee
Because the annihilation and creation operators $d$, $d^\dagger$ do not depend parametrically on $a$, there is no particle production while the Universe expands. Additionally, we consider a coupling between those three fields. This interaction term generates transitions between different Fock states of the fields. The total matter Hamiltonian decomposes into a free part and an interaction term,   
\be
\hat H_\ma(a) = \hat H_0(a) + \hat H_I(a) . \label{Hdecomp}
\ee
We assume the simplest form for the interaction, that is proportional to the product of field operators $\propto \hat \psi_M \hat \psi_\mu \hat \phi$. Explicitly 
\be
\hat H_I(a) = \frac{g(a)}{2Ga} \left[ d_\mu^{\dagger} d_\phi^{\dagger} d_M + d_M^{\dagger} d_\phi^{\dagger} d_\mu + d_\mu^{\dagger} d_M^{\dagger} d_\phi + d_\mu d_\phi d_M + \textrm{h.c.}\right]. \label{Hint}
\ee
The prefactor is chosen here for latter convenience, and changing it would amount to a redefinition of the coupling constant $g$. In flat space, such an interaction allows the most massive field $\psi_M$ to decay into the smaller mass field $\psi_\mu$, by emitting a conformal photon. When the Universe is expanding, i.e., when $a$ becomes dynamical, transitions that were forbidden occur. The light field can be excited into the heavy field and emit a conformal photon. This model can be interpreted as follow. The conformal field is at equilibrium with the expanding Universe. The two massive fields undergo transitions due to the interaction with the conformal field. Exciting the light one into the heavy one is interpreted as absorbing a conformal photon. Therefore, the massive fields play the role of an Unruh-DeWitt detector~\cite{Unruh76,BirrellDavies} probing the state of the radiation field $\phi$. For instance, in a de Sitter background, the massive fields thermalized with the conformal field at the de Sitter temperature. The state is then characterized by the Boltzmann ratio of the mean occupation numbers
\be
\frac{\bar n_M}{\bar n_\mu} = e^{- \beta \Delta M}.
\ee
In App.~\ref{dS_App}, as a warming-up exercise, we present a detailed derivation of this result using the Schrödinger representation for the fields. 

\subsection{Mass hierarchy}
In most realistic cosmological scenarios, the gravitational degrees of freedom are much heavier than the matter degrees of freedom. This simply comes from the fact that the total mass of the Universe is much bigger than that of a single particle contained in it. Additionally, if one wishes to use a gravitational variable as a ``clock'', it is common wisdom that the latter should not ``fluctuate'' too much. This hand waving argument justifies our approach of the problem, namely to treat differently gravitational and matter degrees of freedom. More precisely, we split degrees of freedom into two classes: heavy and light ones. Following the work of~\cite{Massar97,Massar98} we will fruitfully use an analogy between the present problem and that of electronic transitions in molecular collisions. More precisely, if we consider electrons orbiting around a nucleus of position $R$, in the one-dimensional  external potential $V(R)$, the Schrödinger equation at fixed total energy $E_{\rm Tot}$ reads 
\be
\left[ -\frac1{2M} \p_R^2 + V(R) + \hat H_\el(R) - E_{\rm Tot} \right] | \Psi(R) \rangle = 0, \label{SchroEq}
\ee 
where $\hat H_\el(R)$ is the Hamiltonian for the electrons and depends parametrically on the nuclear position $R$. By comparing the above equation \eqref{SchroEq} to \eq{Wheeler-DeWittEq}, we see that both problems are closely analogous. In the molecular context, the nucleus is much heavier than the electrons. For this reason, nuclear states are described in a WKB basis, while the electronic ones in an \emph{instantaneous} basis, i.e., by considering $R$ as an fixed parameter. Because $R$ is in fact dynamical, \eq{SchroEq} induces transitions among electronic states. In~\cite{Delos72I}, by a using a WKB in $R$-space, the authors investigated the validity of the classical trajectory approximation for the nucleus. This would correspond to the background metric approximation in quantum cosmology. However, the molecular problem suffers from the same limitations in the vicinity of a turning point. In this context, a turning point is a common situation, and it is hard to believe that the main conclusions should be affected. In~\cite{Delos72II}, they extended the results in such configurations by using a WKB basis in momentum space. In this work we shall adopt a similar method to investigate matter transitions in the vicinity of a bounce.
\bigskip 

Even though our method is inspired by the above discussion of hierarchy of scales, we shall be very careful with the various approximations employed. More precisely, we will derive the effective equation from the \emph{full} (minisuperspace) Wheeler-DeWitt equation so as to identify precisely what is neglected to reach a ``unitary regime''. This will turn our qualitative statements of ``gravitational degrees of freedom are heavier than matter degrees of freedom'' into precise mathematical conditions. Before proceeding to this derivation, we briefly discuss the specific features of a bounce, and why the Vilenkin-Massar-Parentani interpretation fails in this context.

\subsection{$a$ as a time variable and the Vilenkin-Massar-Parentani interpretation}

In the work of Vilenkin-Massar-Parentani~\cite{Vilenkin88,Massar97,Massar98}, the authors investigate the notion of unitarity when $a$ is chosen as a time variable. They proposed to exploit the conserved current, i.e., the Wronskian of the Wheeler-DeWitt equation $i(\Psi^* \p_a \Psi - \Psi \p_a \Psi^*)$. Since this quantity is not positive definite, unitarity fails to be an exact concept. However, when the time variable is heavy enough, positive and negative sectors decouple, and the current furnishes a reliable notion of scalar product and associated probability interpretation. Note that the results of~\cite{Massar97,Massar98} generalize other approaches involving the Born-Oppenheimer approximation~\cite{Wudka89,Venturi90,Kiefer93,Bertoni96} as no ``tight wave packet'' hypothesis is needed to obtain a unitary evolution. More precisely, it is not assumed that the spread on energy of the matter sector is small. By doing so, the unitary regime clearly appears before a notion of background metric can be defined. In the next section, the same hierarchy of regimes shall be discussed with care. 

In this work, we investigate what happens near turning points of the scale factors. Semiclassically, it is clear why $a$ cannot be a reliable time variable, since it is no longer a monotonic function of time. At the level of the Wheeler-DeWitt equation, it is slightly trickier, since one cannot relate $a$ to any time coordinate. However, it is a well-known feature that WKB approximation schemes fail around turning points, and hence $a$ will no longer be heavy enough in the sense of Vilenkin. A last way to see the failure of the Vilenkin-Massar-Parentani interpretation near turning points is that the current vanishes. Indeed, on the other ``side'' of the turning point (in the region classically forbidden for $a$), the wave function decays exponentially. Because of this, the corresponding solution of the Wheeler-DeWitt equation is real. Therefore, one cannot consistently separate its positive and negative parts and make sense of a probabilistic interpretation. Very similar conclusions have been reached in other approaches of the problem of time as in~\cite{Bojowald10,Bojowald10b,Hohn11}, showing that such a break down of $a$ as a clock is intrinsic, and due to its lack of inertia. Hence, near a turning point, a different degree of freedom must be used to parametrize the Wheeler-DeWitt equation and the physics encoded in it. For this, we follow the same logic as Vilenkin-Massar-Parentani, that is we choose a time variable and assume it to be heavy enough to use a gradient expansion. The appropriate procedure, as is generically the case near a turning point, is to switch to the momentum representation, i.e., to use $\P$ as a time variable. 

\section{Matter states transitions in momentum representation}
In this section, we derive and discuss the dynamics of the Wheeler-DeWitt equation when using $\P$ to parametrize the rate of matter transitions. As we shall see, our derivation is in fact more general than the model described in the previous section. We shall only take profit of the decomposition of \eq{Hdecomp}. We define in general terms the free eigenbasis by 
\be
\hat H_0(a) |n \rangle = E_n(a) |n\rangle. \label{diab_basis}
\ee
This basis is independent of $a$, and is orthonormal, i.e. 
\be
\langle n | m \rangle = \delta_{n,m}.
\ee
When applied to the model of \Sec{content_Sec}, $n$ stands for the triplet $(n_M,n_\mu, n_\phi)$, and $E_n(a) = n_M M + n_\mu \mu + n_\phi k/a$. In full generality, the choice of a basis for the matter states is a delicate issue. There are generally two canonical choices, whether we diagonalize the free Hamiltonian $\hat H_0$ (\emph{diabatic basis}) or the total Hamiltonian $\hat H_0 + \hat H_I$ (\emph{adiabatic basis}). The first case, which is the one we retained, is appropriate for weak couplings, while the other one would be adapted for strong couplings~\cite{Delos79}. Notice also than \eq{Hint} implies that diagonal elements of $\hat H_I$ vanish. If it were not the case, it is straightforward to see that those can be absorbed in a redefinition of $\hat H_0$, i.e., diagonal elements only contribute a shift of the energy levels. 

\subsection{Free case}
\label{pWKB_Sec}

When neglecting the interaction term (i.e., $g = 0$), the particle numbers of each field are conserved. In other words, the total wave function can be projected on an element $|n\rangle$ of the eigenbasis of $H_0$. \eq{Wheeler-DeWittEq} then reduces to 
\be
\left[- G^2 \p_a^2 - V_\eff(a) - 2Ga E_n(a) \right] \Psi_n(a) = 0, 
\ee
where $\Psi_n(a) = \langle n | \Psi(a) \rangle$. To lighten the notations, we define the reduced potentials at fixed $n$, $W_n(a) = - G^{-2}(V_{\rm eff}(a) + 2Ga E_n(a))$. We now turn into momentum representation, taking the Fourier transform of the wave function 
\be
\tilde \Psi(\P) = \int \Psi(a) e^{-i a \P} \frac{da}{\sqrt{2\pi}}.
\ee
At this level, it is worth noting that the assumption that this Fourier transform is well-defined imposes a \emph{boundary condition} on $a$, as done e.g. in~\cite{Kiefer88}. By doing so, we automatically discard the solution growing exponentially in $a$ on the forbidden side of the turning point. Such an assumption seems unavoidable if one wishes to recover unitarity and the background field approximation in some regime, since working with the $a$-time simply fails. Therefore one is forced to switch to the momentum representation. At fixed $n$, the wave function in $\P$ obeys the equation 
\be
\left[\P^2 + W_n(i\p_{\P})\right] \tilde \Psi_n(\P) = 0. \label{PEqfixedn}
\ee
To solve this equation approximately, we look for solutions under the \emph{ansatz} 
\be
\tilde \Psi(\P) = A_n(\P) e^{-i \int^{\P} a_n(\P') d\P'}.
\ee
This turns \eq{PEqfixedn} into 
\be
\left[\P^2 + W_n\Big(a_n(\P) + i\p_{\P}\Big)\right] A_n(\P) = 0. \label{unexpandWKB}
\ee
At this level, we perform a gradient expansion, that is, we assume that the phase of the wave function varies much faster than its amplitude. For this we use a Taylor expansion of $W_n$ in the ``variable'' $\p_{\P}$ (as done e.g. in~\cite{Delos72II}). One must be careful and use formulas for functions of non-commuting arguments. Here, we use the first order expansion 
\be
W_n\Big(a_n(\P) + i\p_{\P}\Big) = W_n(a_n(\P)) + W_n'(a_n(\P)) i\p_{\P} + \frac i2 W_n''(a_n(\P)) a_n'(\P) + O\left(\p_{\P}^2 \right). 
\ee
Notice that at first order in $\p_{\P}$, this is the only ordering of $W_n'(a_n(\P))$ and $\p_{\P}$ that preserve the hermiticity of the operator. Plugging this into \eq{unexpandWKB}, we obtain 
\be
\left[\P^2 +W_n(a_n(\P)) + W_n'(a_n(\P)) i\p_{\P} + \frac i2 W_n''(a_n(\P)) a_n'(\P) \right] A_n(\P) = 0. \label{WKBexpan}
\ee
When sorting these terms in gradients, we see that the first order one vanishes if and only if $\P^2 +W_n(a_n(\P)) = 0$. This is nothing else than the classical Hamilton-Jacobi equation at fixed matter energy 
\be
G^2 \P^2 -V_\eff(a_n(\P)) - E_n(a_n(\P)) = 0. \label{piHJ}
\ee
The second order part of \eq{WKBexpan} gives the amplitude $A_n$ 
\be
A_n(\P) = \left|W_n'(\P)\right|^{-1/2}. \label{slowA}
\ee
From this we obtain the approximate solution of the Wheeler-DeWitt equation 
\be
\chi_n(\P) = \frac{e^{-i \int a_n(\P')d\P'}}{\sqrt{|W_n'(\P)|}}. \label{nWKB} 
\ee
In this derivation, we carefully avoided speaking of a ``WKB approximation'', even though our solution was obtained the same way, i.e. by a gradient expansion. The reason is that this approximation does {\it not} amount to treating the gravitational degrees of freedom as classical. More precisely, we obtained WKB solution for {\it each} matter state $|n\rangle$. Each of these solutions are associated with a different semiclassical trajectory $a_n(\P)$. A general solution of the Wheeler-DeWitt equation is now obtained by superposition 
\be
\tilde \Psi(\P) = \sum_n C_n \chi_n(\P) |n\rangle. \label{freeg0decomp}
\ee
Such a {\it multi}-WKB approximation is very common when studying systems with light internal degrees of freedom, for instance in the standard discussion of the Stern and Gerlach experiment~\cite{Gottfried}. However, in this case it goes further because the semiclassical trajectories $a_n(\P)$ do not possess any intrinsic time coordinate. At the level of \eq{freeg0decomp}, the notion of a single background has no meaning. To recover the evolution in a background metric, we must perform an expansion of $a_n(\P)$ around the mean value of the matter energy $\bar E = E_{\bar n}$, i.e., 
\bsub \label{BFA_cond}
\bea
a_n(\P) &\sim& a_{\bar n}(\P) + \p_E a_n(\P) (E_n - \bar E), \label{BFA_a} \\
W_n(\P) &\sim& W_{\bar n}(\P). 
\eea \esub 
Only then, one can identify the background solution $\dot{\bar p}_a(t) = (\p_E a)^{-1}$. This follows from the Hamilton-Jacobi relations\footnote{We notice that the sign convention seems unusual. In molecular physics, this is due to the fact that we vary with respect to the internal energy $E_\el$ and not with respect to the total energy $E_{\rm Tot}$. In quantum cosmology, it is slightly trickier, and might be related to the issue of the arrow of time. Here, we simply do not want to enter into such delicate issues, and chose the sign so as to recover the correct background metric approximation.}. Note that the approximation \eqref{BFA_cond} amounts to considering ``tight wave packets'', something usually done to define the so-called ``WKB time''~\cite{Isham92,Kiefer93,Bertoni96}. Under assumptions \eqref{BFA_cond}, the wave function then reads 
\be
\tilde \Psi(\P) = \frac{e^{-i \int \bar a(\P')d\P' + i \bar E t}}{\sqrt{W_{\bar n}'(\P)}} \sum_n c_n e^{-i E_n t} |n\rangle.
\ee
We now see that the solution is a superposition of matter states evolving in the background time $t$, multiplied by a WKB wave function for the gravitational part. We recall that the aim of this last discussion, was to justify that the {\it multi}-WKB approximation of \eq{nWKB}, still encode quantum gravitational effects. This will be even clearer in the next section, where matter transitions are included. 

\subsection{Transitions due to interactions}
\label{ptime_Sec}
We now include the interaction Hamiltonian $\hat H_I$ of \eq{Hint}. Because of this interaction, the particle numbers are no longer conserved. Hence, in the decomposition of \eq{freeg0decomp} the coefficients $C_n$ may now depend on the time variable $\P$. 
\be
\tilde \Psi(\P) = \sum C_n(\P) \chi_n(\P) |n\rangle. 
\ee
Using it, the Wheeler-DeWitt equation reads 
\be
0 = \sum_n \left[\P^2 + W_n(i\p_{\P}) + \hat V_I(i\p_{\P}) \right] C_n(\P) \chi_n(\P) |n\rangle,
\ee
where $\hat V_I(a) = 2G^{-1}a \hat H_I(a)$ is the ``dressed'' interaction. We now use the same gradient expansion as in \Sec{pWKB_Sec} and we obtain 
\be
0 = \sum_n \chi_n(\P) \left[W_n'(a_n(\P)) i\p_{\P} + \hat V_I(a_n(\P)) \right] C_n(\P) |n\rangle.
\ee
As before, the first order terms disappear because $a_n$ obey the Hamilton-Jacobi equation \eqref{piHJ}. In addition, we have worked perturbatively in both $g$ (interaction term) and gradients. Hence, at this level of approximation $\hat V_I(i\p_{\P}) \sim \hat V_I(a_n(\P))$. Projecting the equation above with $\langle n |$ and isolating $C_n$, we get 
\be
-i W_n'(\P) \chi_n(\P) \p_{\P} C_n(\P) = \sum_{m \neq n} \langle n | \hat V_I(a_m(\P)) | m \rangle \chi_m(\P) C_m(\P).
\ee
Using the expression for $\chi$ in \eq{nWKB}, we finally obtain 
\be
-i \sign(W_n') \p_{\P} C_n(\P) = \sum_{m \neq n} \frac{\langle n | \hat V_I(a_m(\P)) | m \rangle}{\sqrt{\left| W_n'(\P)W_m'(\P) \right|}} e^{-i \int \left( a_n(\P') - a_m(\P') \right) d\P'} C_m(\P). \label{PWKB_dynEq}
\ee
To obtain this equation, we have made 2 main approximations 
\ben
\item The momentum WKB approximation at fixed matter state $n$. We recall that it is \emph{not} equivalent to treat gravity classically, since here we consider {\it many} WKB wave function and thus superposition of different semiclassical trajectories. 
\item That the interaction terms $\langle n| \hat V_I | m \rangle$ are fairly approximated by their {\it zeroth} order in $\P$-gradients. This is a legitimate approximation if one aims at computing transition amplitudes perturbatively.
\een
However, \eq{PWKB_dynEq} is \emph{not} yet unitary in $\P$ evolution. To obtain a unitary equation, one must make 2 extra assumptions 
\ben[resume]
\item The dressed interaction $\hat V_I(a_m(\P))$ must be parametrically independent of the matter state $|m \rangle$. 
\label{Hint_item}
\een
At this level, there is a natural conserved inner product, which reads in this basis 
\be
\sum_n \sign(W_n') |C_n(\P)|^2 = \textrm{const}. \label{Inner_prod}
\ee
This is not yet a scalar product, as it is not positive definite. Hence, it fails to provide a consistent probability interpretation. In order to recover unitarity and the usual probability interpretation, a last condition must be fulfilled: 
\ben[resume]
\item The semiclassical ``monotonicities'' $W_n'(a_n(\P))$ must all have the same sign. We recall that Hamilton's equations give $\dot{p}_a = - \p_a H_{\rm Tot} = W_n'$. Therefore, classically, the sign of $W_n'$ determines whether $\P$ is increasing or decreasing with respect to time. Note that the absolute sign has no significance here, as it is purely conventional, what matters is the relative sign.
\label{ForceItem}
\een
Probabilities are now well-defined and their sum is conserved, i.e., 
\be
\sum_n |C_n(\P)|^2 = \textrm{const}. \label{Scal_prod}
\ee 
This last condition shows that non-unitarities are due to the coupling between semiclassical trajectories with a \emph{different} monotonicity of $\P$ with respect to time. Interestingly, this is also what was found in~\cite{Massar97,Massar98} when using $a$ as a time variable. In this latter case, non-unitary transitions were due to the coupling between expanding and contracting solutions. This conclusion is generalized when studying the evolution in momentum representation. Semiclassical trajectories evolving in the other time-direction affect the conserved inner product by contributing with a minus sign, as is clearly obtained in \eq{Inner_prod}. A unitary evolution is then recovered when these two classes of semiclassical trajectories decouple. In addition, the variable $\P$ must be have enough inertia so that the WKB approximation of \eq{nWKB} is valid. It can be shown (see App.\ref{PWKB_App}) that a general criterion for this is that for all relevant values of $\P$, we have 
\be
\left|\frac{W_n''^2 a_n'}{W_n'^2}\right| \ll 1. \label{PWKB_cond}
\ee
In particular, the $\P$-WKB approximation becomes increasingly good in the vicinity of a turning point, where $a_n'(\P) = 0$. This regime is orthogonal to what was found using $a$ as a time. In the latter case, unitarity seems highly violated around a turning point, but in momentum time, unitarity emerges to a very good approximation.  

Our analysis therefore indicates that unitarity violations are induced by the failure of our time variable to be a ``good clock''. The educated guess we would have made for this, i.e. $\P$ is monotonic in time and do not fluctuate too much ($\P$-WKB is valid) is exactly what is found from the Wheeler-DeWitt equation, even though the notion of background time is not yet well-defined. Therefore, unitarity seems to be a notion intrinsically prior to that of 
a background space-time. Because of this, time in the conditional sense (i.e., using $\P$) does not yet correspond to its ``Schrödinger'' notion~\cite{Isham92}, i.e., it cannot be seen as a coordinate of some underlying space-time. As we now detail, the background notion of time emerges when considering transitions at linear order in energy change.

\subsection{Recovering the background field approximation}
\label{BFA_Sec}
It is instructive to discuss the background metric limit again, from the dynamics of matter states. Starting from \eq{PWKB_dynEq}, we recover a notion of background geometry by considering energy changes at first order only. Hence, using the expansion of \eq{BFA_cond}, the dynamics of transitions is given by 
\be
-i \sign(W_{\bar n}') \p_{\P} C_n(\P) = \sum_{m \neq n} \frac{\langle n | \hat V_I(\bar a(\P)) | m \rangle}{\left| W_{\bar n}'(\P) \right|} e^{-i \int \left( E_n(\P') - E_m(\P') \right) \p_E a_{\bar n}(\P') d\P'} C_m(\P). 
\label{fromPWKBtoBFA}
\ee
On the right-hand side, in the exponential factor, we recognize $\p_E a_{\bar n}(\P') d\P' = d\P/\dot \P = dt$, where we identify $t$ with the background time. More precisely, it is when we identify the energy in the expansion \eqref{BFA_a} that we are choosing a time coordinate to parametrize the metric. Time is therefore defined as the conjugate variable to energy. In our model, by choosing $E_{n}(a) = n_M M + n_\mu \mu + n_\phi k/a$, we define the comoving time. If we identify the energy as $\tilde E_{n}(a) = a n_M M + a n_\mu \mu + k n_\phi$, we would obtain $\bar a$ in terms of the conformal time. The same is true when identifying the interaction Hamiltonian $\hat H_I$ starting form the dressed interaction $\hat V_I$. Here we use the co-moving time, that is, $\hat V_I(a) = 2G^{-1}a \hat H_I(a)$. Doing so, the prefactor on the left-hand side of \eq{fromPWKBtoBFA} is exactly what is needed to change $\p_{\P}$ into the background time derivative $\p_t$. This is a direct consequence of the Hamilton-Jacobi equation and the expansions of \eq{BFA_cond}. Doing so, we obtain  
\be
-i \p_t C_n(t) = \sum_{m \neq n} \langle n | \hat H_I(\bar a(t)) | m \rangle e^{-i \int \left( E_n(t') - E_m(t') \right) dt'} C_m(t). 
\label{BFA_dynEq}
\ee
This is precisely what we would obtain starting from the Schrödinger equation $i\p_t |\psi \rangle = \hat H_\ma |\psi \rangle$ in the background metric $\bar a(t)$. Notice that the term $\sign(W_{\bar n}')$ of \eq{PWKB_dynEq} plays a crucial role to recover the background field approximation.

As a last remark, we would like to point out that the expansion of \eq{BFA_a} is slightly subtle, as $a(\P)$ is not a parametric function of the total matter energy.  However, by considering infinitesimal variation $\delta a$, $\delta E$ to the Hamilton-Jacobi equation \eqref{piHJ}, one sees that the first order expansion of \eq{BFA_a} is well defined. This shows that the background metric limit is ``universal'', but next-to-leading order corrections may depend on the specific matter content. 

\section{Matter transitions in the vicinity of the bounce}
\label{transition_Sec}
\subsection{Phase space trajectories}
\begin{figure}[!ht]
\begin{center} 
\includegraphics[width=0.8\columnwidth]{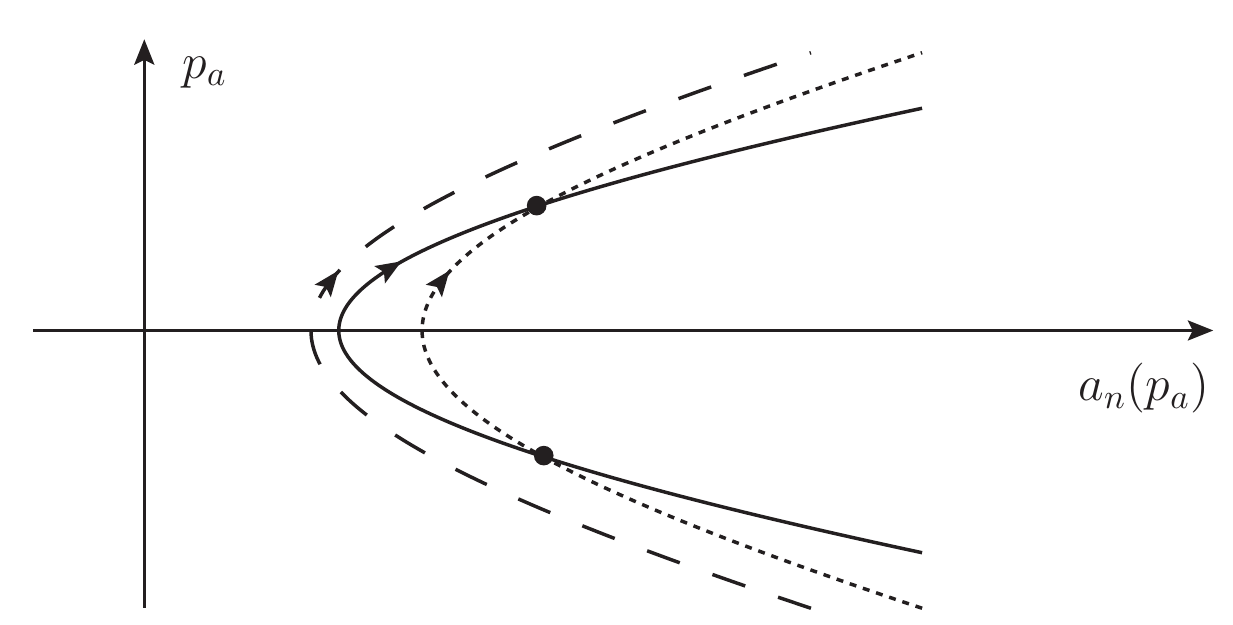}
\end{center}
\caption{Semiclassical trajectories $a_n(\P)$ for different values of $n$. The continuous line is $n_i$. Dashed lines are two possible $n_f$, with either two or zero crossing points.  
}
\label{Lvlcrossing_fig} 
\end{figure}

We now describe how matter transitions that occur in a close vicinity of a bounce. To do so, we approximate the effective potential by a linear function around the bounce value of $a = a_b$, i.e., 
\be
V_\eff(a) = \kappa (a - a_b) . \label{Veff_bounce}
\ee 
When implementing this assumption in the Hamilton-Jacobi equation, we obtain 
\be
G^2 \P^2 - \kappa (a - a_b) - 2Ga (n_M M + n_\mu \mu + n_\phi k/a) = 0. \label{HJ_bounce}
\ee
We see that having $a_b > 0$, i.e., a bounce does occur, amounts to having a \emph{negative} density of radiation. Such a violation of positive-energy conditions is to expect, since otherwise, the usual singularity theorems state that an initial singularity necessarily occurs~\cite{Wald}. \eq{Veff_bounce} would also be obtained by linearizing the effective potential around the recollapse of a closed Universe. From \eq{HJ_bounce}, we deduce the semiclassical scale factor functions $a_n(\P)$. As we saw in the preceding section, they govern the transitions of matter states through \eq{PWKB_dynEq}, and differ for each matter state $| n \rangle = | n_\mu, n_M, n_\phi \rangle$. Using \eq{HJ_bounce} and expressing $a$ as a function of $\P$, we see that they take the simple form 
\be
a_n(\P)  = a_{b,n} + R_n \P^2 .
\ee
The presence of matter and radiation dresses the value for the turning point and the ``acceleration rates'' $R_n$. From \eq{HJ_bounce}, it follows that 
\bsub \label{Rabn}
\bea
a_{b,n} &=& \frac{ \kappa a_b - 2G n_\phi k}{\kappa + 2G n_M M + 2G n_\mu \mu} , \label{abn} \\
R_n &=& \frac{G^2}{\kappa + 2G n_M M + 2G n_\mu \mu} . \label{Rn} 
\eea \esub
From the preceding expressions, we see that the notion of a single background metric is still absent, since Eqs.~\eqref{abn}, \eqref{Rn} do not linearly depend on matter energy, and hence \eq{BFA_cond} is not exact. To understand a transition $|n_i \rangle \to |n_f \rangle$, we first look at the corresponding semiclassical trajectories. Here we see that depending on the relative sign of $\Delta a_b = a_{b,n_f} - a_{b,n_i}$ and $\Delta R = R_{n_f} - R_{n_i}$, there is either two level crossings, or none, see Fig.~\ref{Lvlcrossing_fig}. As it is generically the case for Landau-Zener types of transitions, when no crossing occurs, the transition probability is exponentially small. On the other hand, when two crossing occurs, the probability to make a transition will be appreciable, and given by a sum of two interfering terms. This interference is govern by the phase shift accumulated during the propagation between the two crossing points. As we shall now see, this general argument gives the correct answer only when the two crossing points are well separated.  

\subsection{Matter transitions}
To further simplify, we now assume that the coupling constant $g$ of \eq{Hint} is independent of $a$. Together with the linear effective potential of \eq{Veff_bounce}, the Wheeler-DeWitt equation \eqref{Wheeler-DeWittEq} becomes first order in $\p_{\P}$. Therefore, the $\P$-WKB approximation is exact, in agreement with \eq{PWKB_cond}. Lastly, the various $W_n' = -R_n^{-1}$ all possess the same sign. Note that because of the minus sign, the momentum is decreasing with time for each semiclassical solution involved. Therefore, initial state will be specified for $\P \to +\infty$ and final state for $\P \to -\infty$. 

We now compute the probability amplitude to make a transition from an initial state $|n_i \rangle$ to a final state $|n_f \rangle$. At ``early times'', well before the bounce, the matter  state has definite numbers of particle $| n_i \rangle = | n_\mu, n_M, n_\phi \rangle$. The instantaneous amplitudes evolve in decreasing values of $\P$, and the final state becomes a superposition of states $\sum C_m |m \rangle$. The probability to measure the system in the final state $|n_f \rangle$ will then be given by the square of the amplitude $C_{n_f}$ at ``late time''. At first order in perturbation theory, one can replace the $C_n$ on the right-hand-side of \eq{PWKB_dynEq} by their initial value, i.e. $C_m = \delta_{m, n_i}$. The transition is thus driven by a single matrix element  
\be
\M = \sqrt{|R_{n_i} R_{n_f}|} \langle n_f | \hat V_I | n_i \rangle. 
\ee
Integrating \eq{PWKB_dynEq} between $-\infty$ and $+\infty$, we express the final amplitude as the integral 
\bsub \bea
C_{n_f}(-\infty) &=& i \M \int e^{-i \int \left( a_{n_i}(\P') - a_{n_f}(\P') \right) d\P'} d\P ,\\
&=& i \M \int e^{i (\Delta a_b \P + \Delta R \P^3/3)} d\P .
\eea \esub
We recognize here the integral representation of an Airy function~\cite{Olver}. Taking care of the sign of $\Delta R$, the probability amplitude of the considered transition is given by 
\be \label{transit_rate}
A_{n_i \to n_f} = 2i\pi \frac{\M}{|\Delta R|^{1/3}} \Ai\left(\sign(\Delta R) \frac{\Delta a_b}{|\Delta R|^{1/3}} \right). 
\ee
When the argument of the Airy function is large, the result is simple to interpret. As we anticipated, when $\Delta a_b$ and $\Delta R$ have the same sign, the transition probability is exponentially suppressed, 
\be
A_{n_i \to n_f} \sim i \frac{\M \sqrt \pi}{|\Delta R|^{1/12}|\Delta a_b|^{1/4}} e^{-\frac23 |\Delta a_b|^{3/2}|\Delta R|^{-1/2}}. 
\ee
This is due to the absence of crossing of the semiclassical trajectories. On the other hand, when they have a different sign, the asymptotic of the Airy function the product of an amplitude and an oscillating term, 
\be
A_{n_i \to n_f} \sim \frac{\M \sqrt \pi}{|\Delta R|^{1/12}|\Delta a_b|^{1/4}} \left( e^{i\frac23 |\Delta a_b|^{3/2}|\Delta R|^{-1/2} + i\frac\pi4} - e^{-i\frac23 |\Delta a_b|^{3/2}|\Delta R|^{-1/2} - i\frac\pi4} \right). \label{interf_transition}
\ee
The amplitude is the (appreciable) probability of making a transition at each crossing. These two events interfere with a phase shift due to the propagation from one crossing to the other. Note that in more realistic models, this interference is likely to disappear due to interactions with other degrees of freedom and the decoherence effect they induce. 

In fact, both results for large Airy argument could have been obtained by studying the transitions with $a$ as a time parameter. On the contrary, when $\Delta a_b \lesssim |\Delta R|^{1/3}$, the amplitude is essentially governed by $\M/|\Delta R|^{1/3}$ and barely depends on $\Delta a_b$ 
\be
A_{n_i \to n_f} \sim 2i\pi \frac{\M}{|\Delta R|^{1/3}} \Ai(0). 
\ee
This result cannot be obtained using $a$ as the time variable. This shows the relevance of using $\P$ as a time. One cannot simply compute the evolution of matter states using $a$ away from the turning point, and complete the solution using a connection formula of the type $C_n^{\rm expansion} \to e^{i\pi/2} C_n^{\rm contraction}$, i.e., the standard WKB completion around turning points~\cite{Gottfried,Delos72I}. Such a connection formula would give \eq{interf_transition}. However, the transitions satisfying $\Delta a_b \lesssim |\Delta R|^{1/3}$ occur in the vicinity of the bounce, and can only be obtained in momentum time.

Additionally, \eq{transit_rate} shows that semiclassical trajectories that substantially differ, i.e., when $\Delta R$ and $\Delta a_b$ are large, decouple, since $A_{n_i \to n_f} \to 0$. Nevertheless, when semiclassical trajectories get closer, but $R_n$ and $a_{b,n}$ of \eq{Rabn} are still non-linear in energy, the dynamics of matter transitions cannot be obtained in the background metric approximation. This shows the necessity of the effective unitary regime (i.e., \eq{PWKB_dynEq} under assumptions \ref{Hint_item} and \ref{ForceItem}) to go beyond the background metric approximation.

\section{Conclusion}
The aim of this paper was to investigate the notion of unitarity in quantum cosmology near a bounce. Around a turning point, matter transitions seem highly non-unitary, and the Vilenkin current fails to provide us with a consistent probabilistic interpretation. On the contrary, when we parametrize the evolution using the conjugate momentum $\P$, we find a well-defined hierarchy of three regimes. The first one, without any approximation, does not possess any conserved scalar product, and a consistent probability interpretation seems to be absent. 

In a second regime, one recovers a unitary evolution, while the notion of a background metric is still absent. To reach this regime, several conditions must be satisfied. First, the variable $\P$ must not ``fluctuate too much'', so that the momentum WKB approximation is valid, see \eq{PWKB_cond}. Second, semiclassical trajectories corresponding to different monoticities in time must decouple, as they contribute to the conserved product with a different sign, see \eq{Inner_prod}. We underline that this is quite similar to what was found using $a$ as time~\cite{Massar97,Massar98}, although in an orthogonal regime (in the sense of \eq{PWKB_cond}). There, these two conditions are in fact the same: the departure from the $a$-WKB approximation is exactly due to the coupling between contracting and expanding solutions. Lastly, the interaction Hamiltonian should not parametrically depend on the state of matter fields. It is worth pointing out that this approximate unitary regime goes \emph{beyond} the ``WKB time'' proposition (see e.g.~\cite{Isham92,Kiefer13} and references therein). Indeed, at this level, the gravitational degrees of freedom are not described by a single WKB wave. Hence, the notion of background metric, and with it, the ``Schrödinger'' notion of time, is still absent. 

In the third regime, the notion of background metric emerges and one recovers quantum field theory in curved space-time with the mean backreaction. To do so, the changes in matter energy must be small enough in order to be treated at \emph{first} order. As pointed out in~\cite{Parentani96}, this is similar to the emergence of temperature when a small system is coupled to a thermal bath. The variation of energy of the small system must be negligible with respect to that of the total system. 

We finally point out that even though the obtained notion of unitarity and probability interpretation is approximate, it leads to highly non-trivial physical predictions. In the intermediate regime, the backreaction of matter on gravity is is still quantum in nature, that is, it goes beyond the mean field approximation. Moreover, as investigated in \Sec{transition_Sec}, if matter transitions occur in a close vicinity of the bounce, the probability amplitude can only be obtained using $\P$ as time. This shows that switching to momentum representation is not only necessary at the conceptual level, it also leads to non-trivial physical predictions. 

It would be interesting to extend the present formalism to larger classes of metric, like Bianchi models. More generally, including several heavy degrees of freedom, treated as we treated the scale factor, should display new interesting features. Interestingly, the corresponding problem exists in molecular physics~\cite{Miller72}. This would be especially relevant for approaches using the inflaton field as a time variable, as for instance in loop quantum cosmology. Additionally, this might lead to non-trivial new effects such as non-zero Berry phases~\cite{Balbinot90}. We hope to investigate these new cases in a forthcoming future. 

\acknowledgements
I am deeply grateful to Renaud Parentani and Serge Massar for many valuable discussions and the time they took to explain their previous works. I would also like to thank warmly Sylvain Carrozza, Marc Geiller, and Philipp Höhn for several fruitful discussions and many useful comments on the final version of this work.

\appendix
\section{Gibbons-Hawking effect in Schrödinger representation}
\label{dS_App}
In this section we consider a fixed background metric given by the de Sitter space. Using \eq{FLRW} the space-time metric is fully characterized by 
\be
a(t) = a_0 e^{H t}, \label{dSa}
\ee
where $t$ is the comoving time, which means that we work in the gauge $N=1$. The Hubble rate is related to the cosmological constant by $H = \Lam/\sqrt 3$. When the background metric is well-defined, matter fields evolve following the Schrödinger equation in a curved space-time, i.e., 
\be
i\p_t |\psi \rangle = \hat H_\ma(a(t)) |\psi \rangle. \label{Schro}
\ee
Because the Hamiltonian is time dependent, there is no notion of stationary states. Using the free basis, as in \eq{diab_basis}, the time-dependence of the Schrödinger equation induces transitions among matter states. Using such a basis with time-dependent coefficients, the dynamics for the instantaneous amplitudes is given by 
\be
-i \p_t C_n(t) = \sum_{m \neq n} \langle n | \hat H_I(a(t)) | m \rangle e^{-i \int \left( E_n(t') - E_m(t')\right) dt'} C_m(t).
\ee
Note that this equation is simply a rewriting of \eq{Schro} in the chosen basis; it contains no approximation~\cite{Massar98}. We consider the transition corresponding to the light field being spontaneously excited into the heavy field by emitting a conformal photon, i.e., 
\be
\mu \to M + \gamma. \label{mu_decay}
\ee
At first order in perturbation theory, we only need to consider the matrix element $\langle n_i | \hat H_I | n_f \rangle$. The probability amplitude for this transition, given by the asymptotic value of $C_{n_f}$, is given by the integral  
\be
A_{\mu \to M} = i \int_{-\infty}^{+\infty} \langle n_i | \hat H_I | n_f \rangle e^{i \int \left(E_{n_f}(t') - E_{n_i}(t') \right) dt'} dt. \label{BFA_1st_transit}
\ee
By evaluating this integral with a saddle point approximation, we can anticipate the various cases. The main contribution will come from the saddle point $t^*$, solution of $E_{n_f}(t) = E_{n_i}(t)$. If there is a level crossing, that is, $t^*$ is real, then the saddle point mainly contributes as an oscillating phase, and the transition probability is appreciable. This can be interpreted by the fact that at the time $t^*$, making a transition costs no energy. On the contrary, when there is no level crossing, the solution $t^*$ lies in the complex plane. Then the saddle point contributes as a decaying exponential, and the transition amplitude is correspondingly small. This comes from the fact that conservation of energy prevents the transition from happening ``classically''. Only through a tunnel effect can the transition be achieved. Therefore, the transition of \eq{mu_decay} would vanish in flat space. Indeed, the difference in energy, 
\be
E_{n_f}(t) - E_{n_i}(t) = M - \mu + \frac{k}{a_0} e^{-H t} 
\ee
never vanishes. In this case we can compute the transition amplitude analytically. To simplify, we assume that the matrix element $\langle n_i | \hat H_I | n_f \rangle$ is independent of $a$, and hence, of $t$. \eq{BFA_1st_transit} then gives 
\be
A_{\mu \to M} = i \langle n_i | \hat H_I | n_f \rangle \int_{-\infty}^{+\infty} e^{i\left( (M - \mu)t - \frac{k}{a_0 H} e^{-H t}\right)} dt.
\ee
By making the change of variable $X = k e^{-H t}/a_0$, we obtain the representation of an Euler $\Gamma$ function. Explicitly, 
\be
A_{\mu \to M} = i \frac{\langle n_i | \hat H_I | n_f \rangle}{H} \left(\frac{k}{a_0}\right)^{i \frac{M-\mu}{H}} e^{-\frac{(M - \mu)\pi}{2H}} \Gamma\left(-i\frac{M - \mu}{H}\right).
\ee
If one further integrates over all values of $k$, one recovers the time dependence of the usual golden rule~\cite{Massar98}. The complete expression is not essential, as everything can be understood when comparing this result to the amplitude of the transition 
\be
M \to \mu + \gamma. 
\ee
Here, up to the phase, the only factor that is affected is the exponential $e^{-\frac{(M - \mu)\pi}{2H}}$. This gives a simple relation between the transition probabilities 
\be
|A_{\mu \to M}|^2 = e^{-\frac{2\pi (M - \mu)}{H}} |A_{M \to \mu}|^2. \label{dSratio}
\ee
Therefore, using the standard Einstein argument, we see that the massive fields shortly reach an equilibrium state. The ratio of the occupation numbers at equilibrium gives the Boltzmann law 
\be
\frac{n_M}{n_\mu} = e^{- \beta_{\rm dS} (M - \mu)},
\ee
and one recovers the de Sitter temperature $\beta_{\rm dS}^{-1} = T_{\rm dS} = 2\pi/H$ (in units where $k_B = 1$). Note additionally, that taking into account a time-dependence of the matrix element $\langle n_i | \hat H_I | n_f \rangle$, will \emph{not} affect this last result. Indeed, the probability amplitudes will be affected, but the ratio, i.e., \eq{dSratio} is maintained, as all what is needed is the analytic continuation of the exponential behavior of the scale factor in \eq{dSa}. This is very similar to what happens when studying the Unruh or the Hawking effect~\cite{Primer}.

\section{Adiabatic approximation}
\label{Adiab_App}
In this section, we justify the assumption of ``heavy fields'' made in \Sec{content_Sec} and derive the validity condition to neglect non-adiabatic transitions. To shorten the discussion, we work within the background metric approximation, with a line element given by \eq{FLRW}. The action of a (minimally coupled) massive field $\varphi$ of vanishing momentum reads
\be
\mathcal S_\ma = \frac12 \int \left( \dot \varphi^2 - \mu^2 \varphi^2 \right) a^3 dt. 
\ee
For simplicity, we directly wrote the action in the gauge $N=1$, that is with the comoving time $t$. To absorb the $a^3$ factor, we redefine the field  
\be
\psi = a^{3/2} \varphi .
\ee
Putting this into the action, one gets
\be
\mathcal S_\ma = \frac12 \int \left( \dot \psi^2 - \mu^2 \psi^2 - \frac{3\dot a}{a} \dot \psi \psi + \frac{9\dot a^2}{a^2} \psi^2 \right) dt. 
\ee 
After integration by part to eliminate the linear term in $\dot \psi$, we obtain 
\be
\mathcal S_\ma = \frac12 \int \left( \dot \psi^2 - \mu_{\rm eff}^2(t) \psi^2 \right) dt, 
\ee
with
\be
\mu_{\rm eff}^2(t) = \mu^2 - \frac{3(\dot a^2 + 2 a \ddot a)}{4a^2}.
\ee
In this rewriting, pair creation effects are entirely due to the time dependance of $\mu_{\rm eff}$. If $\p_t \mu_{\rm eff}/ \mu_{\rm eff}^2 \ll 1$, there is no particle creation, and the action is that of a time-independent harmonic oscillator. However, one might want to make a stronger approximation, where not only particles are not spontaneously created, but also the mass keeps its undressed value $\mu$ and is not affected by the expansion. For this we need $|\mu_{\rm eff}^2 - \mu^2|/\mu^2 \ll 1$, and if derivatives of $a$ are well sorted, it reduces to the criterion 
\be
\left|\frac{\dot a^2}{\mu^2 a^2}\right| \ll 1.
\ee
With a similar computation, we obtain the condition of adiabaticity for a (minimally coupled) massless field of momentum $k$ mentioned in the footnote \ref{massless_ftn} of \Sec{content_Sec}. 

\section{The momentum WKB approximation}
\label{PWKB_App}
In this section we sketch the argument leading to the criterion of \eq{PWKB_cond} for the validity of the WKB approximation in momentum space. For this we reinvestigate the derivation of \Sec{pWKB_Sec} but keeping the next-to-leading order corrections. To simplify, we drop the $n$-dependence and without loss of generality assume $W' > 0$. We start by assuming a wave function of the form 
\be
\tilde \Psi(\P) = A(\P) e^{-i \int^{\P} a(\P') d\P'} (1+\eps(\P)),
\ee
where $A(\P)$ is the slowly varying amplitude found in \eq{slowA} and $\eps(\P)$ the correction to the WKB wave function. Putting this into the Wheeler-DeWitt equation \eqref{PEqfixedn}, we obtain 
\be
\left[\P^2 + W\Big(a(\P) + i\p_{\P}\Big)\right] A(\P)(1+\eps(\P)) = 0. \label{amplWKBeq}
\ee
To go beyond \Sec{pWKB_Sec}, we expand $W$ to second order in $i\p_{\P}$. The second order term in $i\p_{\P}$ is a combination of  
\be
\frac12 i\p_{\P} W''(a(\P)) i\p_{\P} \qquad \textrm{and} \qquad \frac14 \left((i\p_{\P})^2 W''(a(\P) + W''(a(\P)) (i\p_{\P})^2\right), \label{ordering}
\ee
which are the only orderings compatible with the hermitian character of $\hat W$. 
To shorten the argument, we shall consider only the first. As we argue below, the precise ordering does not affect the final criterion\footnote{In fact, using symmetry arguments, one can show that the correct ordering is half the sum of the two terms in \eq{ordering}.}. In \eq{amplWKBeq}, this term will only act on $A$, not on $\eps$ as we only keep terms up to second order in gradient expansion. Therefore, we see that \eq{amplWKBeq} reduces to 
\be
W'(a(\P)) A(\P) i \eps'(\P) + \frac12 i\p_{\P} W''(a(\P)) i\p_{\P} A(\P) = 0. 
\ee
Just like in \Sec{pWKB_Sec}, the first order term vanishes because $a(\P)$ is solution of the Hamitlon-Jacobi equation, and similarly, the second order term vanishes since $A(\P) = W'(\P)^{-1/2}$. After a tedious but straightforward calculation, we obtain $\eps'(\P)$ expressed by $W$ and its derivatives 
\be
\eps'(\P) = \frac{i}8 \left[\frac{4W' W'' W''' a'^2 + 2 W' W'' a'' - 3 W''^3 a'^2}{W'^3}\right]. \label{epsprime}
\ee
At this level we see that the exact evaluation of WKB corrections is not an easy task. However, by assuming that the potential is smooth enough that its derivatives are well sorted, all terms contribute to the same order. The numerical factors of the various terms above are thus irrelevant, and one can tune them to integrate $\eps'(\P)$ in a simple manner. Note  that the precise combination of \eq{ordering} would in fact change the numerical factor of \eq{epsprime}. We now integrate \eq{epsprime} and obtain the local correction to the momentum WKB approximation 
\be
\eps(\P) \simeq i \left[\frac{W''^2 a'}{W'^2}\right].
\ee
The criterion of \eq{PWKB_cond} is then simply $|\eps| \ll 1$. Note that one can rewrite this condition by deriving the Hamilton-Jacobi equation $\P^2 + W(a(\P)) = 0$ and eliminate $a'$. This gives the alternative condition 
\be
\left|\frac{2W_n''^2 \P}{W_n'^3}\right| \ll 1. 
\ee
This last result agrees with the claim of reference~\cite{Delos72II}.

\bibliographystyle{utphys}
\bibliography{Bibli}

\providecommand{\href}[2]{#2}\begingroup\raggedright\begin{thebibliography}{10}

\bibitem{DeWitt67a}
B.~S. DeWitt, ``{Quantum Theory of Gravity. 1. The Canonical Theory},''
\href{http://dx.doi.org/10.1103/PhysRev.160.1113}{{\em Phys. Rev.} {\bfseries
  160} (1967) 1113--1148}.

\bibitem{UnruhWald89}
W.~G. Unruh and R.~M. Wald, ``{Time and the Interpretation of Canonical Quantum
  Gravity},''
\href{http://dx.doi.org/10.1103/PhysRevD.40.2598}{{\em Phys.Rev.} {\bfseries
  D40} (1989) 2598}.

\bibitem{Isham92}
C.~Isham, ``{Canonical quantum gravity and the problem of time},'' in {\em
  Integrable Systems, Quantum Groups and Quantum Field Theories}, Salamanca
  proceedings.
\newblock 1992.
\newblock
\href{http://arxiv.org/abs/gr-qc/9210011}{{\ttfamily arXiv:gr-qc/9210011
  [gr-qc]}}.
\newblock

\bibitem{CarterHartle}
J.~Hartle, ``Prediction in quantum cosmology,'' in {\em Gravitation in
  Astrophysics}, B.~Carter and J.~Hartle, eds., vol.~156 of {\em Adv. Study
  Inst. Ser. B Phys.}, pp.~pp. 329--360, NATO ASI Series.
\newblock Plenum Press,
1987.
\newblock

\bibitem{Kuchar91}
K.~Kuchar, ``{Time and interpretations of quantum gravity},''
\href{http://dx.doi.org/10.1142/S0218271811019347}{{\em Int. J. Mod. Phys.
  Proc. Suppl.} {\bfseries D 20} (2011) 3--86}.

\bibitem{Brout98}
R.~Brout and R.~Parentani, ``{Time in cosmology},''
  \href{http://dx.doi.org/10.1142/S0218271899000031}{{\em Int. J. Mod. Phys.}
  {\bfseries D 8} (1999) 1--22},
\href{http://arxiv.org/abs/gr-qc/9705072}{{\ttfamily arXiv:gr-qc/9705072
  [gr-qc]}}.

\bibitem{Halliwell84}
J.~Halliwell and S.~Hawking, ``{The Origin of Structure in the Universe},''
\href{http://dx.doi.org/10.1103/PhysRevD.31.1777}{{\em Phys. Rev.} {\bfseries D
  31} (1985) 1777}.

\bibitem{Brout88}
R.~Brout and G.~Venturi, ``{Time in Semiclassical Gravity},''
\href{http://dx.doi.org/10.1103/PhysRevD.39.2436}{{\em Phys. Rev.} {\bfseries D
  39} (1989) 2436}.

\bibitem{Born26}
M.~Born, ``Zur quantenmechanik der sto{\ss}vorg{\"a}nge,''
  \href{http://dx.doi.org/10.1007/BF01397477}{{\em Zeitschrift f{\"u}r Physik A
  Hadrons and Nuclei} {\bfseries 37} no.~12, (1926) 863--867}.

\bibitem{Vilenkin88}
A.~Vilenkin, ``{The Interpretation of the Wave Function of the Universe},''
\href{http://dx.doi.org/10.1103/PhysRevD.39.1116}{{\em Phys. Rev.} {\bfseries D
  39} (1989) 1116}.

\bibitem{Massar97}
S.~Massar and R.~Parentani, ``{Particle creation and nonadiabatic transitions
  in quantum cosmology},''
  \href{http://dx.doi.org/10.1016/S0550-3213(97)00718-9}{{\em Nucl. Phys.}
  {\bfseries B 513} (1998) 375--401},
\href{http://arxiv.org/abs/gr-qc/9706008}{{\ttfamily arXiv:gr-qc/9706008
  [gr-qc]}}.

\bibitem{Massar98}
S.~Massar and R.~Parentani, ``{Unitary and nonunitary evolution in quantum
  cosmology},'' \href{http://dx.doi.org/10.1103/PhysRevD.59.123519}{{\em Phys.
  Rev.} {\bfseries D 59} (1999) 123519},
\href{http://arxiv.org/abs/gr-qc/9812045}{{\ttfamily arXiv:gr-qc/9812045
  [gr-qc]}}.

\bibitem{KeskiVakkuri96}
E.~Keski-Vakkuri and S.~D. Mathur, ``{Quantum gravity and turning points in the
  semiclassical approximation},''
  \href{http://dx.doi.org/10.1103/PhysRevD.54.7391}{{\em Phys. Rev.} {\bfseries
  D 54} (1996) 7391--7406},
\href{http://arxiv.org/abs/gr-qc/9604058}{{\ttfamily arXiv:gr-qc/9604058
  [gr-qc]}}.

\bibitem{Hajicek86}
P.~Hajicek, ``{Origin of Nonunitarity in Quantum Gravity},''
\href{http://dx.doi.org/10.1103/PhysRevD.34.1040}{{\em Phys. Rev.} {\bfseries D
  34} (1986) 1040}.

\bibitem{Marolf94}
D.~Marolf, ``{Almost ideal clocks in quantum cosmology: A Brief derivation of
  time},'' \href{http://dx.doi.org/10.1088/0264-9381/12/10/007}{{\em Class.
  Quant. Grav.} {\bfseries 12} (1995) 2469--2486},
\href{http://arxiv.org/abs/gr-qc/9412016}{{\ttfamily arXiv:gr-qc/9412016
  [gr-qc]}}.

\bibitem{Marolf09}
D.~Marolf, ``{Solving the Problem of Time in Mini-superspace: Measurement of
  Dirac Observables},''
  \href{http://dx.doi.org/10.1103/PhysRevD.79.084016}{{\em Phys. Rev.}
  {\bfseries D 79} (2009) 084016},
\href{http://arxiv.org/abs/0902.1551}{{\ttfamily arXiv:0902.1551 [gr-qc]}}.

\bibitem{Bojowald10}
M.~Bojowald, P.~A. Hoehn, and A.~Tsobanjan, ``{An Effective approach to the
  problem of time},''
  \href{http://dx.doi.org/10.1088/0264-9381/28/3/035006}{{\em Class. Quant.
  Grav.} {\bfseries 28} (2011) 035006},
\href{http://arxiv.org/abs/1009.5953}{{\ttfamily arXiv:1009.5953 [gr-qc]}}.

\bibitem{Bojowald10b}
M.~Bojowald, P.~A. Hohn, and A.~Tsobanjan, ``{Effective approach to the problem
  of time: general features and examples},''
  \href{http://dx.doi.org/10.1103/PhysRevD.83.125023}{{\em Phys. Rev.}
  {\bfseries D 83} (2011) 125023},
\href{http://arxiv.org/abs/1011.3040}{{\ttfamily arXiv:1011.3040 [gr-qc]}}.

\bibitem{Hohn11}
P.~A. Hohn, E.~Kubalova, and A.~Tsobanjan, ``{Effective relational dynamics of
  a nonintegrable cosmological model},''
  \href{http://dx.doi.org/10.1103/PhysRevD.86.065014}{{\em Phys. Rev.}
  {\bfseries D 86} (2012) 065014},
\href{http://arxiv.org/abs/1111.5193}{{\ttfamily arXiv:1111.5193 [gr-qc]}}.

\bibitem{Wald}
R.~Wald, {\em General relativity}.
\newblock University of Chicago press, 1984.

\bibitem{Bojowald12}
M.~Bojowald, ``{Quantum Cosmology: Effective Theory},''
  \href{http://dx.doi.org/10.1088/0264-9381/29/21/213001}{{\em Class. Quant.
  Grav.} {\bfseries 29} (2012) 213001},
\href{http://arxiv.org/abs/1209.3403}{{\ttfamily arXiv:1209.3403 [gr-qc]}}.

\bibitem{Ashtekar11}
A.~Ashtekar and P.~Singh, ``{Loop Quantum Cosmology: A Status Report},''
  \href{http://dx.doi.org/10.1088/0264-9381/28/21/213001}{{\em Class. Quant.
  Grav.} {\bfseries 28} (2011) 213001},
\href{http://arxiv.org/abs/1108.0893}{{\ttfamily arXiv:1108.0893 [gr-qc]}}.

\bibitem{Bekenstein75}
J.~Bekenstein, ``{Nonsingular General Relativistic Cosmologies},''
\href{http://dx.doi.org/10.1103/PhysRevD.11.2072}{{\em Phys. Rev.} {\bfseries D
  11} (1975) 2072--2075}.

\bibitem{Kiefer13}
C.~Kiefer, ``{Conceptual Problems in Quantum Gravity and Quantum Cosmology},''
  \href{http://dx.doi.org/10.1155/2013/509316}{{\em ISRN Math. Phys.}
  {\bfseries 2013} (2013) 509316},
\href{http://arxiv.org/abs/1401.3578}{{\ttfamily arXiv:1401.3578 [gr-qc]}}.

\bibitem{Parentani96}
R.~Parentani, ``{Time dependent Green functions in quantum cosmology},''
  \href{http://dx.doi.org/10.1016/S0550-3213(97)00139-9}{{\em Nucl. Phys.}
  {\bfseries B 492} (1997) 475--500},
\href{http://arxiv.org/abs/gr-qc/9610044}{{\ttfamily arXiv:gr-qc/9610044
  [gr-qc]}}.

\bibitem{Parentani96b}
R.~Parentani, ``{Time dependent perturbation theory in quantum cosmology},''
  \href{http://dx.doi.org/10.1016/S0550-3213(97)00140-5}{{\em Nucl. Phys.}
  {\bfseries B 492} (1997) 501--525},
\href{http://arxiv.org/abs/gr-qc/9610045}{{\ttfamily arXiv:gr-qc/9610045
  [gr-qc]}}.

\bibitem{BirrellDavies}
N.~Birrell and P.~Davies, {\em Quantum fields in curved space}.
\newblock Cambridge University Press, 1984.

\bibitem{Unruh76}
W.~G. Unruh, ``Notes on black-hole evaporation,''
  \href{http://dx.doi.org/10.1103/PhysRevD.14.870}{{\em Phys. Rev.} {\bfseries
  D 14} no.~4, (1976) 870--892}.

\bibitem{Delos72I}
J.~B. Delos, W.~R. Thorson, and S.~K. Knudson, ``Semiclassical theory of
  inelastic collisions. i. classical picture and semiclassical formulation,''
  {\em Phys. Rev.} {\bfseries A 6} no.~2, (1972) 709.

\bibitem{Delos72II}
J.~B. Delos and W.~R. Thorson, ``Semiclassical theory of inelastic collisions.
  ii. momentum-space formulation,'' {\em Phys. Rev.} {\bfseries A 6} no.~2,
  (1972) 720.

\bibitem{Wudka89}
J.~Wudka, ``{A Comment on the Born-Oppenheimer Approximation},''
\href{http://dx.doi.org/10.1103/PhysRevD.41.712}{{\em Phys. Rev.} {\bfseries D
  41} (1990) 712}.

\bibitem{Venturi90}
G.~Venturi, ``{Minisuperspace, matter and time},''
\href{http://dx.doi.org/10.1088/0264-9381/7/6/014}{{\em Class. Quant. Grav.}
  {\bfseries 7} (1990) 1075--1087}.

\bibitem{Bertoni96}
C.~Bertoni, F.~Finelli, and G.~Venturi, ``{The Born-Oppenheimer approach to the
  matter - gravity system and unitarity},''
  \href{http://dx.doi.org/10.1088/0264-9381/13/9/005}{{\em Class. Quant. Grav.}
  {\bfseries 13} (1996) 2375--2384},
\href{http://arxiv.org/abs/gr-qc/9604011}{{\ttfamily arXiv:gr-qc/9604011
  [gr-qc]}}.

\bibitem{Kiefer93}
C.~Kiefer, ``{The Semiclassical approximation to quantum gravity},'' in {\em
  {Canonical gravity-from classical to quantum}}, J.~Ehlers and H.~Friedrich,
  eds.
\newblock 1993.
\newblock
\href{http://arxiv.org/abs/gr-qc/9312015}{{\ttfamily arXiv:gr-qc/9312015
  [gr-qc]}}.
\newblock

\bibitem{Delos79}
J.~Delos and W.~Thorson, ``Diabatic and adiabatic representations for atomic
  collision processes,'' {\em The Journal of Chemical Physics} {\bfseries 70}
  (1979) 1774.

\bibitem{Kiefer88}
C.~Kiefer, ``{Wave Packets in Minisuperspace},''
\href{http://dx.doi.org/10.1103/PhysRevD.38.1761}{{\em Phys. Rev.} {\bfseries D
  38} (1988) 1761}.

\bibitem{Gottfried}
K.~Gottfried and T.~Yan, {\em Quantum mechanics: fundamentals}.
\newblock Springer Verlag, 2nd~ed., 2003.

\bibitem{Olver}
F.~Olver, {\em Asymptotics and special functions}, vol.~15.
\newblock Academic Press New York, 1974.

\bibitem{Miller72}
W.~H. Miller and T.~F. George, ``Semiclassical theory of electronic transitions
  in low energy atomic and molecular collisions involving several nuclear
  degrees of freedom,'' {\em The Journal of Chemical Physics} {\bfseries 56}
  (1972) 5637.

\bibitem{Balbinot90}
R.~Balbinot, A.~Barletta, and G.~Venturi, ``{Matter, quantum gravity, and
  adiabatic phase},''
\href{http://dx.doi.org/10.1103/PhysRevD.41.1848}{{\em Phys. Rev.} {\bfseries D
  41} (1990) 1848--1854}.

\bibitem{Primer}
R.~Brout, S.~Massar, R.~Parentani, and P.~Spindel, ``{A Primer for black hole
  quantum physics},''
  \href{http://dx.doi.org/10.1016/0370-1573(95)00008-5}{{\em Phys. Rept.}
  {\bfseries 260} (1995) 329--454},
\href{http://arxiv.org/abs/0710.4345}{{\ttfamily arXiv:0710.4345 [gr-qc]}}.

\end{thebibliography}\endgroup

\end{document}